# SECURING OUR BLUETOOTH MOBILES FROM INTRUDER ATTACK USING ENHANCED AUTHENTICATION SCHEME AND PLAUSIBLE EXCAHNGE ALGORITHM

Ms.A.Rathika
Assistant Professor
VCET, Erode, India
.

Ms. R.Saranya
Final Year UG Student
VCET, Erode, India

Ms.R.Iswarya
Final Year UG Student
VCET, Erode, India

*ABSTRACT-* **When Bluetooth devices come within the range of another, an electronic conversation takes place to determine whether the devices in range are known or whether one needs to control the other. Most Bluetooth devices do not require any form of user interaction for this to occur. If devices within range are known to one another, the devices automatically form a network-known as a pairing.**

**Authentication addresses the identity of each communicating device. The sender sends an encrypted authentication request frame to the receiver. The receiver sends an encrypted challenge frame back to the sender. Both perform a predefined algorithm. The sender sends its findings back to the receiver, which in turn either allows or denies the connection.**

**There are three different functions for authentication in Bluetooth-E1, E2, and E3. E1 is used when encrypting the authorization challenge-response values.E2 is for generating different link keys.E3 is used when creating the encryption key.**

## *Key words: link key, primitive root, challenge response scheme*

## GENERALIZATION OF INITIALIZATION KEY:

The creation of an initialization key is used when no other keys are present. The key is derived from a random number, a PIN, length of the PIN and a unit's hardware address. The PIN code can either be a factory value or the user can enter a maximum of 16 octets.

## GENERALIZATION OF LINK KEY AND LINK KEY EXCHANGE:

When a link key is established between two units they will use that key for authentication. A link key is 128 bits long and a shared between two or more units, a new link key can be derived whenever to improve security.

Each device creates a random no and encrypts it together with its hardware address and produces initialization key..

## AUTHENTICATION:

The Bluetooth authentication procedure is based on a **challenge-response** scheme. Two devices interacting in an authentication procedure are referred to as the claimant and the verifier. The verifier is the blue tooth device validating the identity of another device. The claimant is the device attempting to prove its identity.

The challenge-response protocol validates devices by verifying the knowledge of a secret key- a Bluetooth link key. The steps in the authentication process are the following:

- Step1: the claimant transmits its 48-bit address (BD_ADD) to the verifier.
- Step2: the verifier transmits a 128-bit random challenge (AU_RAND) to the claimant.
- Step3: the verifier uses the E1 algorithm to compute an authentication response using the address, link key and random challenge as inputs.
- Step4: the claimant returns the computed response SRES, to the verifier.

- Step5: the verifier compares the SRES from the claimant with the SRES that it computes.
- Step 6: if the two 32-bit SRES values are equal, the verifier will continue connection establishment.

.The E1 authentication function used for the validation is based on the **SAFER+** algorithm.

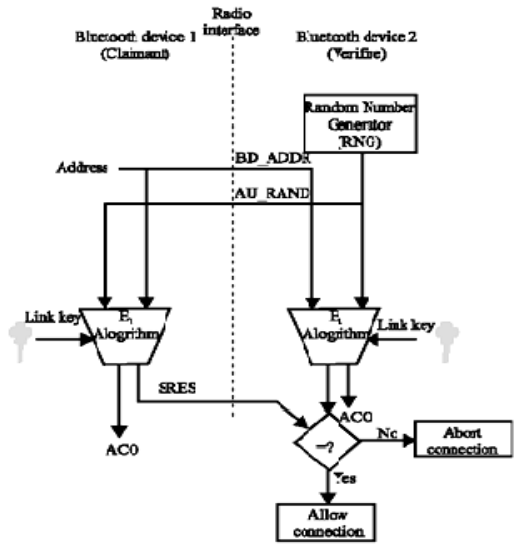

**Fig1: Authentication Process**

**PROBLEM IN THE CURRENT SYSTEM:**

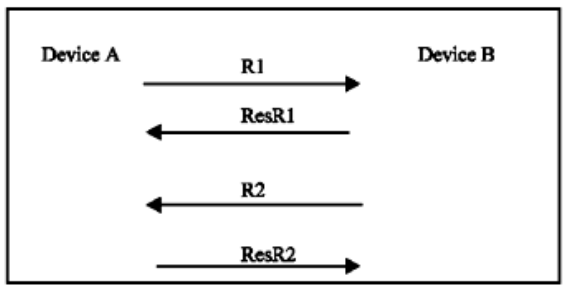

**Fig2: Message in Existing Authentication Process**

When the connection is made between the Bluetooth devices, an intruder device can be there in different ways. An intruder can act as the fake device in the different roles. The fake device can behave as false slave or false master. Similarly the intruder can be a active intruder or passive one. It can continue the connections to the both communicating devices or detach the one end messages sent by intruder C is shown in Fig.3.

In the existing authentication scheme of Bluetooth technology mutual authentication is performed.

First one device sends the random number for authentication to the device second,

Then the second device sends the response and sends another random number for the verification of first device. Then the first device sends the response of random number send b second device. In this way the identification of both the devices is done.

In the above figure .2, device A sends a random number R1 to device B for authentication of device B. Then the device B sends the random number R2 and the device B sends the ResR2 to device B

**Behavior of intruder C in existing scheme:**

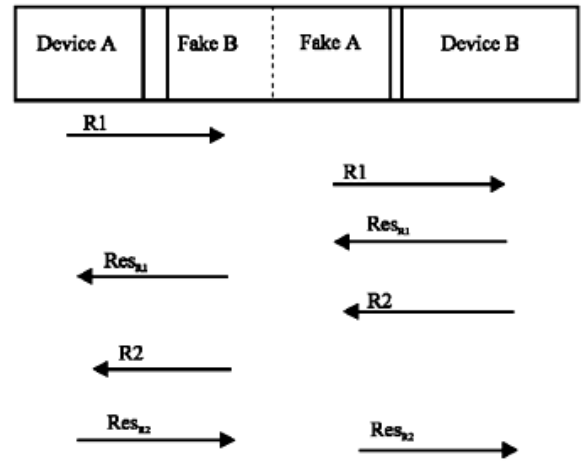

**Fig3:Message in Existing Authentication Process by the intruder**

Suppose an intruder C wants to make connection in between the both devices A and B.

- Device A sends random number R1 to the fake device B.
- Fake device B now behaves as fake device B.
- Now the device B sends the response Res R1 to fake device A.
- Intruder's C sends the authentication random number R2 to fake device A.

- The intruder C sends the random number R2 to device A.
- The device A sends the Res R2 to fake device.
- The fake device now sends the same response Resr2 to device.

Hence in this way the intruder device makes the connection with the devices A and b.

**IMPROVED AUTHENTICATION METHOD:**

In the authentication scheme, mutual authentication is performed exclusively between master and slave. First, one is authenticated with the AU_RAND (challenge) and SRES (response) exchange. Then the other is

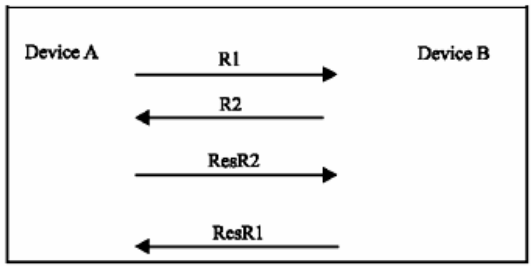

Fig. 4: Messages in new Authentication process

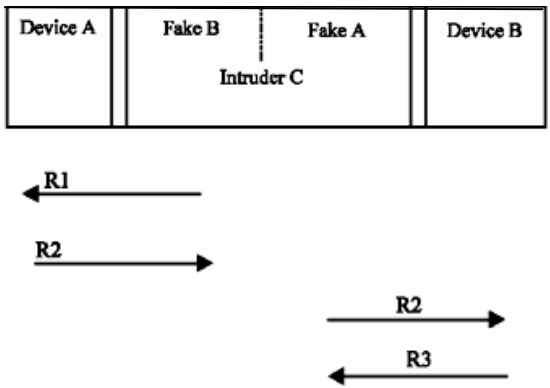

Authenticated again using a challenge response mechanism. We propose to change this authentication message exchanges in a form such that first both parties exchange their authentication random value and claimant does not sends its response before getting the response from the verifier. In this method, the attacker cannot obtain SRES vale from the other party. Since the attacker acts as a verifier in both piconets, its authentication challenge is responded with another authentication challenge from the genuine entities.

With the improved authentication method, if messages exchanged in a nested form such that first both parties exchange their random values and claimant does not sends its response from the verifier. The messages are shown as below:

Now there are two case in this authentication procedure:

➢ When the request for connection is generated from the intruder device C to device A.

➢ When the request for connection is made from the device A to device C.

**Case 1: request from c to A:**

In this case when the intruder C will initiate the connection establishment procedure with device A.

- The fake device sends the random number R1 to device A.
- The device A does not sends the response for R1.
- It sends the another random number R2 to fake device B for authentication and waits for the response for R2.

Suppose the fake device is trying to get the response from device B. It sends the same random number to devices B.

The device does not sends the response of R2,firstly it verifies the fake device A and sends the one another random number R3 waits for the response of R3 from fake device.

Hence in this case the attacker cant' involve itself into the devices A and B.

**Case 2: Request from A to C:** in this case when the request is made from device A. The following messages are exchanged between the devices A, C and B

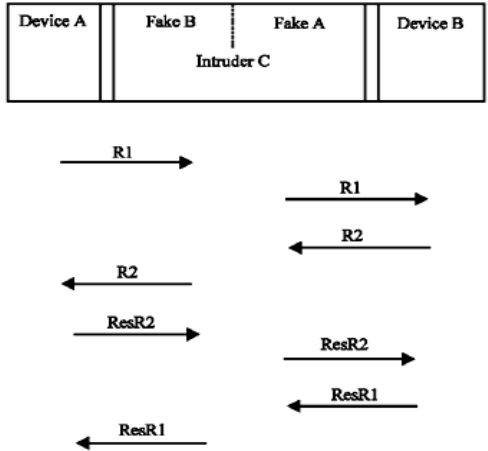

Fig. 6:  Messages in improved Authentication process by intruder (Case 2)

First the authentication random number R1 is sent by device A.

- The device C sends the same random number R1 to device B.
- The device C does not send the response of R1,it sends the another random number R2 to fake device A for authentication.
- The device C transmits the random number R2 to device A and waits for the response of R2 from A.
- The device A gives the response 'ResR2' of R2.
- The device C gives theResR2 to device b.
- The device B sends the ResR1in response to the number R1 to device C.
- The resR1 is sent as it is to device A by the device C.

Hence the connection is made between the devices A and C and C and B, but this is only possible only when the request is initiated by the device A and simultaneously there is a connection between the device A and B.

Integrity is maintained, but the confidentiality is disturbed.

## IMPLEMENTING MORE SECURITY FOR RANDOM NUMBER EXCHANGE:

The scheme implemented in case 2 was insecure obviously. So by implementing the encryption in key exchange we can have a new improved authenticated system with more security to avoid the above said intruder attack.

## PLAUSIBLE EXCHANGE

### DIFFIE-HELLMAN KEY EXCHANGE:

Device A selects a random number R1 such that  R1 < P a prime number which is having α as its primitive root and calculates S1= $a^{R1}$ mod p. Similarly device Selects R2 such that R2 < p and calculates S2 = $a^{R2}$ mod p. Each side keeps the R value Private and S values Public.

Device A computes K = $S2^{R1}$ mod p and B computes K = $s1^{R2}$ mod p so the produce identical results.

So even though the intruder knows the P, S and α value they can predict the S1 and S2 values. But they couldn't predict the original random number.

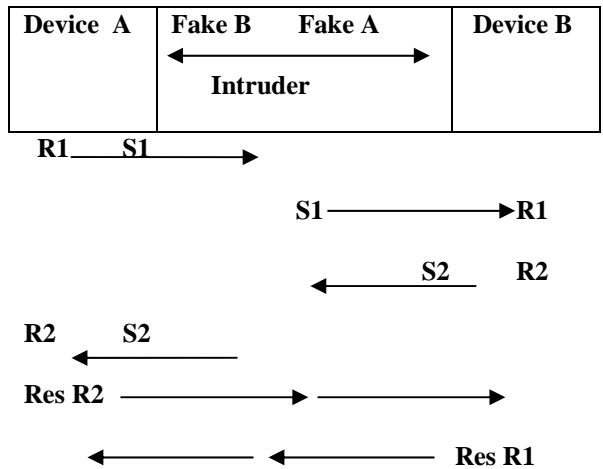

## PROCEDURE TO EXCHANGE KEYS:

**Device A:**  S1= $a^{R1}$ mod p

**Device B:** S2 = $a^{R2}$ mod p

**Device A:**  K = $s2^{R1}$ mod p

**Device B:**  K = $s1^{R2}$ mod p

**ADVANTAGES:**

- This algorithm uses the discrete logarithmic function, which is an irreversible function and it can not be easily decrypted.

- The secret integers S1 and S2 are discarded at the end of the session. Therefore, Diffie-Hellman key exchange by itself trivially achieves perfect forward secrecy because no long-term private keying material exists to be disclosed.

**CONCLUSIONS:**

While Bluetooth has several nice features, it fails to be a secure replacement of wires. As we have shown that Bluetooth is susceptible to the attacks by intruders independent of security mechanisms. If an unknown device wants to make connections or request for a service. The proper authentication is followed by authorization and encryption, but authentication process should be such that unknown device would not get response of any random number until and unless it will give response to the random number which it wants to make the connections.

If we give the provision that not any single slave will response until it verifies the identity of other device and another method is that one device can estimate the delay by observing the response time given by the verifier, so in this way we an check the identity of the device and can improve the security.

**Author's profile**

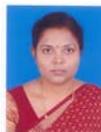

Ms.A.Rathika is now working as an Assistant professor at Velalar College of Engineering and Technology, Erode, Tamil Nadu. She has completed her ME at Anna University, Coimbatore and has more than six years of teaching experience. She has published two papers in journals and also presented many papers in national and international conferences. Her field of interest is Network Security.

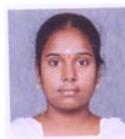

Ms. R. Saranya is currently pursuing her final year B.Tech IT programme in Velalar College of Engineering and Technology, Erode, Tamil Nadu. She has presented many papers in national conference and technical symposium held at various colleges and universities and won laurels. She has more awareness on cryptography and network security which is her field of interest.

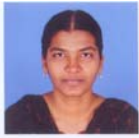

Ms. R. Iswarya is doing her final year B.Tech IT programme in Velalar College of Engineering and Technology, Erode, Tamil Nadu. She has participated and presented many technical papers in national level technical symposium and conferences and gained laurels. Her area of interest is Ethical Hacking and Information Security.